\documentclass[floatfix,twocolumn]{revtex4}
\usepackage{times}
\usepackage{amssymb,amsmath,graphicx}
\usepackage[usenames]{color}

\newcommand{\cPP}{cisPP}
\newcommand{\tPP}{transPP}

\begin{document}

\title{FMO3-LCMO study of electron transfer coupling matrix element and pathway: 
Application to hole transfer between two triptophanes 
through cis- and trans-polyproline-linker systems}

\author{Hirotaka Kitoh-Nishioka\footnote{Present address: 
Center for Computational Sciences, University of Tsukuba, 
E-mail: hkito@ccs.tsukuba.ac.jp} 
\hspace*{0.1em} 
and Koji Ando\footnote{E-mail: ando@kuchem.kyoto-u.ac.jp} }
\affiliation{
Department of Chemistry, Graduate School of Science, Kyoto University, 
Sakyo-ku, Kyoto 606-8502, Japan
}

\begin{abstract}
    The linear-combination of fragment molecular orbitals with three-body correction (FMO3-LCMO)
    is examined for electron transfer (ET) coupling matrix elements and ET pathway analysis,
    with application to hole transfer between two triptophanes bridged by cis- and trans-polyproline
    linker conformations.
    A projection to the 
    minimal-valence-plus-core 
    FMO space was found to give sufficient accuracy
    with significant reduction of computational cost
    while avoiding the problem of linear dependence of FMOs stemming from 
    involvement of bond detached atoms.
\end{abstract}

\maketitle

\section{Introduction}

Long-distance electron transfer (ET) plays an essential role in biological
energy conversion \cite{Moser1992,Dutton1994,Winkler1999,GrayWinkler2005,FarverPecht2011}.
Representative are those in photosynthetic
reaction centers in which photon energy is converted to electrochemical
energy via series of ETs through redox centers embedded in transmembrane
protein.
A simple but fundamental question open to microscopic investigation 
is how the protein environment
is involved in the ET:
the protein structure could be involved passively by
simply holding the redox centers at appropriate spatial configuration,
or actively by providing intermediate virtual states for superexchange
ET mechanism \cite{McConnell1961}.

To address this and related questions, quantum mechanical investigation based
on realistic molecular structure is essential. However, first-principles
treatment of electrons in large biomolecular systems is still a formidable
task. In this regard, fragment-based approaches, such
as the fragment molecular orbital (FMO) \cite{Kitaura1999,Fedorov2007,Tanaka2014}, 
Divide-and-Conquer \cite{Yang95,Akama2009}, 
and many others \cite{Li2010,Aoki2012,Gordon2012,Raghavachari2015}
appear promising.

In a series of papers, we have reported calculations of ET coupling matrix element
and ET pathways \cite{Nishioka2011,Kitoh-Nishioka2012,Nishioka15} 
from the linear-combination of FMOs (FMO-LCMO) \cite{Tsuneyuki2009}
with the two-body correction (FMO2).
The method was found to give accurate ET couplings over 
four orders of magnitude along the ET distance
when the bond detached atoms (BDA) were not involved
or when the minimal atomic basis set was used \cite{Nishioka2011}.
Nevertheless, 
the problem of degraded MO energies caused by BDAs
with atomic basis sets larger than minimal set,
that had been already pointed out in Ref. \cite{Tsuneyuki2009},
carried over to the ET analysis. 
Recently, however,
Kobori et al. have found notable remedy 
of MO energies with the three-body correction (FMO3) \cite{Kobori2013}.
Following this, 
we study in this work how the FMO3 correction affects the ET
coupling energy and ET pathway analysis.

We also examine selection of the FMO space.
In Ref. \cite{Kobori2013}, 
in order to remove linear dependence of basis functions 
associated with BDAs, 
a canonical
transformation of Hamiltonian matrix,
\begin{equation}
\tilde{H}=U^{\dagger}HU ,
\label{eq:UHU}
\end{equation}
where the matrix $U$ diagonalizes overlap matrix,
was employed.
However, this transformation often mixes the FMOs
in unwanted ways 
for the ET pathway analysis, particularly for studying inter-fragment
tunneling current. 
We found that the problem can be evaded by 
a projection to restricted FMO space
instead of the canonical transformation of Eq. (\ref{eq:UHU}).
For instance, with FMO-LC(VC)MO scheme, 
which restricts
to the minimal-valence (V) plus core (C) MO space, 
the smallest eigenvalue
of overlap matrix for systems studied in this work
was 0.225, which is large enough to regard the FMO space linearly independent \cite{SzaboOstlund}.

For numerical demonstration,
we examine hole transfer between two triptophane (Trp) residues bridged
by helical polyproline oligopeptide,
which serves a good model of long-distance ETs observed in
metal-derivatized oligoprolines \cite{Isied1992, Ogawa1993A,Ogawa1993B}.
Previous theoretical works \cite{Wallrapp2009, Wallrapp2010}
have employed the same model systems 
to study the effects of solvent and bridge-conformation dynamics on
the electronic coupling. 

Section \ref{sec:theory} outlines the theoretical framework.
Section \ref{sec:comput} describes the computational details.
Applications to 
hole transfer between two Trp molecules
bridged by
polyproline linker systems
are discussed in Sec. \ref{sec:results}.
Section \ref{sec:concl} concludes.

\section{Theory}
\label{sec:theory}

\subsection{FMO-LCMO and projection to restricted FMO space}

Here we outline the FMO-LCMO method \cite{Tsuneyuki2009,Kobori2013}
to explain the present projection scheme to the restricted FMO space.

The FMO-LCMO Hamiltonian up to the three-body FMO3 correction 
is described as
\begin{equation}
H_\textrm{total}^\textrm{FMO3}=H_\textrm{total}^\textrm{FMO1}
+\Delta H_\textrm{total}^\textrm{FMO2}+\Delta H_\textrm{total}^\textrm{FMO3},
\end{equation}
in which $H_\textrm{total}^\textrm{FMO1}$ consists of monomer Fock matrices,
and $\Delta H_\textrm{total}^\textrm{FMO2}$ and $\Delta H_\textrm{total}^\textrm{FMO3}$ 
are the two- and three-body corrections.
The intra-fragment block of
$\Delta H_\textrm{total}^\textrm{FMO2}$,
between FMOs $\phi^{I}_p$ and $\phi^{I}_q$ in the same fragment $I$, is given by
\begin{equation}
(\Delta H_\textrm{total}^\textrm{FMO2})_{I_{p},I_{q}}
=\sum_{J\ne I}\{(H_{I\leftarrow I\!J})_{I_{p},I_{q}}-(H_{I\leftarrow I})_{I_{p},I_{q}}\} ,
\label{eq:FMO2II}
\end{equation}
whereas the inter-fragment block is
\begin{equation}
(\Delta H_\textrm{total}^\textrm{FMO2})_{I_{p},J_{q}}
=(H_{I\!J\leftarrow I\!J})_{I_{p},J_{q}} .
\label{eq:FMO2IJ}
\end{equation}
The terms in the right-hand-side of Eqs. (\ref{eq:FMO2II}) and (\ref{eq:FMO2IJ})
will be defined in Eqs. (\ref{eq:HII})--(\ref{eq:HIJX}).

The FMO3 correction to the intra- and inter-fragment blocks are 
\begin{multline}
(\Delta H_\textrm{total}^\textrm{FMO3})_{I_{p},I_{q}} 
=\sum_{J<K}\sum_{J,K\ne I}\{(H_{I\leftarrow I\!JK\!})_{I_{p},I_{q}} \\
-(H_{I\leftarrow I\!J})_{I_{p},I_{q}}
-(H_{I\leftarrow I\!K})_{I_{p},I_{q}}+(H_{I\leftarrow I})_{I_{p},I_{q}}\}
\label{eq:FMO3II}
\end{multline}
and
\begin{equation}
(\Delta H_\textrm{total}^\textrm{FMO3})_{I_{p},J_{q}}
=\sum_{K\ne I,J}\{(H_{I\!J \leftarrow I\!J\!K})_{I_{p},J_{q}}
-(H_{I\!J \leftarrow I\!J})_{I_{p},J_{q}}\}.
\label{eq:FMO3IJ}
\end{equation}

As noted in Introduction, the canonical transformation of Eq. (\ref{eq:UHU}) is
not desired for the ET pathway analysis. 
We thus simply limit the number
of FMOs 
and project the Hamiltonian matrix to this set.
We denote this \lq\lq restricted
FMO (rFMO)\rq\rq\ space. Therefore, the terms in the right-hand-side of 
Eqs. (\ref{eq:FMO2II})--(\ref{eq:FMO3IJ}) 
in the selected rFMO space $\{ \phi^{I}_{p} \}$ 
are
defined as follows. With the fragment monomer, dimer, and trimer represented
by $X=I,I\!J$, and $I\!J\!K$, the intra-fragment blocks are defined by
\begin{equation}
(H_{I\leftarrow I})_{I_{p},I_{q}}=\varepsilon_{I_{p}}\delta_{I_{p},I_{q}}
\label{eq:HII}
\end{equation}
and 
\begin{equation}
(H_{I\leftarrow X})_{I_{p},I_{q}}
=\sum_{X_{r}}\langle\phi^{I}_{p}|\phi^{X}_{r}\rangle
\varepsilon^{X}_{r}
\langle\phi^{X}_{r}|\phi^{I}_{q}\rangle ,
\label{eq:HIX}
\end{equation}
whereas the inter-fragment blocks are
\begin{equation}
(H_{I\!J\leftarrow X})_{I_{p},J_{q}}=\sum_{X_{r}}
\langle\phi^{I}_{p}|\phi^{X}_{r}\rangle
\varepsilon^{X}_{r}
\langle\phi^{X}_{r}|\phi^{J}_{q}\rangle .
\label{eq:HIJX}
\end{equation}
In the summation over $X_{r}$ in the right-hand-side, all the dimer
and trimer FMOs are taken,
except the spurious ones stemming from the BDAs.

\subsection{ET coupling and pathway analysis}

To calculate the ET coupling matrix element $T_\textrm{DA}$,
we employ two methods;
generalized Mulliken-Hush (GMH) \cite{Cave1996} and 
bridge Green function (BGF) \cite{Skourtis1999,Regan1999,Stuchebrukhov2003}.

The GMH method scales the donor-acceptor MO energy splitting
$\Delta \varepsilon_\textrm{DA}$ 
by a formula
\begin{equation}
T_\textrm{DA} = 
\frac{|\mu_\textrm{DA}|\; \Delta \varepsilon_\textrm{DA}}
{\sqrt{(\mu_\textrm{D} - \mu_\textrm{A})^2 + 4|\mu_\textrm{DA}|^2}} ,
\label{eq:GMH}
\end{equation}
in which $\mu_\textrm{D}$, $\mu_\textrm{A}$ 
and $\mu_\textrm{DA}$ are the diagonal and off-diagonal dipole matrix elements.
It assumes that 
the Hamiltonian 
and dipole matrix elements scale similarly
for states involved in ETs.
Despite its simplicity, the GMH formula (\ref{eq:GMH}) has been successfully
applied to a number of ET reactions.

The BGF method has been
also demonstrated to give accurate and robust results 
with
\begin{multline}
T_\textrm{DA}=H_{\phi_\textrm{D},\phi_\textrm{A}}^\textrm{direct}
+\sum_{I,J}^{N}\sum_{I_{p},J_{q}}\!{'}
(E_\textrm{tun}S_{\phi_\textrm{D},I_{p}}-H_{\phi_\textrm{D},I_{p}}) \\
\times
G^\textrm{B}(E_\textrm{tun})_{I_{p},J_{q}}(E_\textrm{tun}S_{J_{q},\phi_\textrm{A}}-H_{J_{q},\phi_\textrm{A}}) ,
\end{multline}
in which the sums over $I_p$ and $J_q$ exclude donor and acceptor MOs, $\phi_\textrm{D}$ and $\phi_\textrm{A}$.
The first term in the right-hand-side is the direct coupling between 
$\phi_\textrm{D}$ and $\phi_\textrm{A}$.
$S$ is the overlap matrix.
$G^\textrm{B}(E)$ is the bridge Green function defined as
\begin{equation}
    G^\textrm{B}(E)=(E S_{QQ} - H_{QQ})^{-1} , 
\label{GBE}
\end{equation}
in which $Q$ is the projection operator to the MO space
external to the donor-acceptor MOs.
The electron tunneling energy $E_\textrm{tun}$ 
is naturally defined as the average of
donor-acceptor orbital energies,
\(
E_\textrm{tun} = \left( \varepsilon_\textrm{D} + \varepsilon_\textrm{A} \right) / 2
\) .

In the tunneling current analysis \cite{Stuchebrukhov2003,Stuchebrukhov1996},
the ET coupling $T_\textrm{DA}$ is expressed as a sum of
tunneling current ${\cal J}_{I_p, J_q}$ 
between basis FMOs $\{ \phi^{I}_{p} \}$ \cite{Nishioka2011,Kitoh-Nishioka2012}, 
\begin{equation}
T_\textrm{DA} = \hbar
\sum_{I \in \Omega_\textrm{D}}
\sum_{J \notin \Omega_\textrm{D}}
{\cal J}_{I,J} ,\label{eq:tda}
\end{equation}
\begin{equation}
    {\cal J}_{I,J} = 
\sum_{I_p}
\sum_{J_q}
{\cal J}_{I_p, J_q} ,
\end{equation}
in which the summation over $I_p$ and $J_q$ are 
over the FMOs within fragments $I$ and $J$,
and $\Omega_\textrm{D}$ denotes the spatial region assigned to the donor molecule.
The inter-orbital current ${\cal J}_{I_p, J_q}$ is computed from
the electronic Hamiltonian and overlap matrices and
the coefficients of FMO-LCMO,
$\{ C^\textrm{i}_{I_p} \}$ and $\{ C^\textrm{f}_{I_p} \}$,
that represent the mixing of bridge FMOs to 
the donor and acceptor FMOs, $\phi_\textrm{D}$ and $\phi_\textrm{A}$,
in the initial (\textrm{i}) and final (\textrm{f}) states, 
$\psi^\textrm{i}$ and $\psi^\textrm{f}$,
\begin{equation}
    | \psi^\textrm{i} \rangle = 
    C^\textrm{i}_\textrm{D} | \phi_\textrm{D} \rangle +
\sum_I^N
\sum_{I_p}
C^\textrm{i}_{I_p} | \phi^I_{p} \rangle ,
\end{equation}
\begin{equation}
    | \psi^\textrm{f} \rangle = 
    C^\textrm{f}_\textrm{A} | \phi_\textrm{A} \rangle +
\sum_I^N
\sum_{I_p}
C^\textrm{f}_{I_p} | \phi^I_{p} \rangle ,
\end{equation}
\begin{equation}
    {\cal J}_{I_p, J_q} = \frac{1}{\hbar}
\left( H_{I_p, J_q} - E_{\rm tun} S_{I_p, J_q} \right)
\left( C^\textrm{i}_{I_p} C^\textrm{f}_{J_q} - C^\textrm{f}_{I_p} C^\textrm{i}_{J_q} \right) .
\end{equation}
These are thus computed straightforwardly from 
the FMO-LCMO method.
The normalized inter-fragment tunneling current is defined by
\begin{equation}
    {\cal K}_{I,J}=\hbar {\cal J}_{I,J}/T_\textrm{DA} ,
\end{equation}
which satisfies
\begin{equation}
\sum_{I\in\Omega_\textrm{D},J\notin\Omega_\textrm{D}}{\cal K}_{I,J}=1.
\end{equation}

\section{Computation}
\label{sec:comput}

\subsection{Polyproline linker conformation}
For the purpose of benchmarking 
the FMO3-LCMO calculations,
we consider
proline-trimer bridged systems, Trp-(Pro)$_{3}$-Trp, 
with two types of helix structure of polyproline (PP) linker, 
one with cis-configurations (\cPP)
and the other with trans-configurations (\tPP)
of peptide bonds.
The former and latter proline trimers
have the backbone dihedral angles $(\varphi,\psi)$ of approximately 
$(-75^\circ,150^\circ)$ and $(-75^\circ,160^\circ)$, respectively.
The \cPP\ and \tPP\ structures are schematically drawn in Fig. 1.
Apparently the \tPP\ is more stretched.

\begin{figure}
\includegraphics[width=0.25\textwidth]{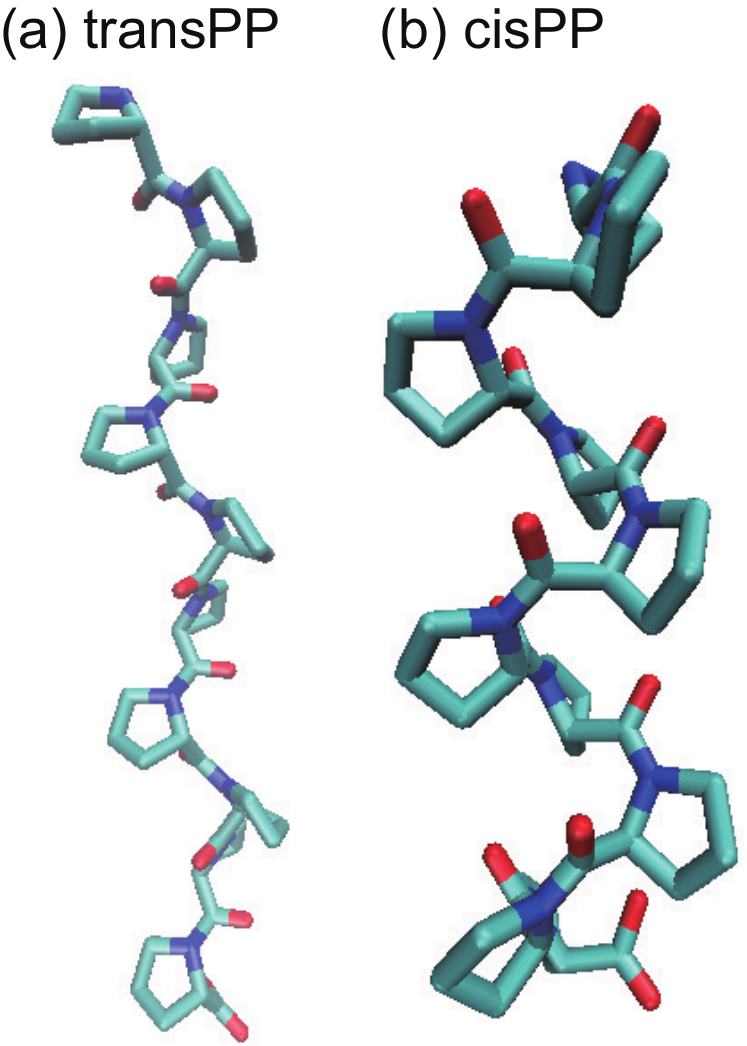}
\caption{
    Schematic drawings of the helix structures of polyproline linker with (a) trans- and (b) cis-configurations.
}
\end{figure}

Starting from typical conformations
of \cPP\ and \tPP, 
local strains 
were removed by geometry optimization
at the B3LYP/6-31G(d)-D3 level, with the D3 version of
Grimme's dispersion correction with the original D3 damping function \cite{Grimme2010}.
We used Gaussian09 program \cite{Gaussian09} for the geometry optimization.
The resultant molecular structures are displayed in 
Fig. S1 and Table S1 of the Supplementary Material.

\subsection{FMO calculation}

In the FMO calculation, the Trp-(Pro)$_{3}$-Trp chain was divided into
six fragments as designated in Fig. 2,
in which P1-P3 are the three prolines and MW denotes the Main-chain of Trp (W).
The $\alpha$-carbon atoms
were treated as BDAs. 
The HOMOs of Trp
fragments were taken as the donor and acceptor MOs of hole transfer. 
The electronic
coupling and tunneling current from the FMO2-LCMO and FMO3-LCMO methods
were compared to a reference calculation with 
the RHF Hamiltonian of the entire system of six fragments projected to the FMO space.
We denote this last scheme FMO6-LCMO with the Hamiltonian matrix
\begin{equation}
(H_\textrm{total}^\textrm{FMO6})_{I_{p},J_{q}}=\sum_{a}
\langle\phi^{I}_{p}|\psi_{a}\rangle\varepsilon_{a}\langle\psi_{a}|\phi^{J}_{q}\rangle ,
\end{equation}
where $\phi^{I}_{p},\phi^{I}_{q}$ are MOs in the rFMO space
and $\psi_{a}$ and $\varepsilon_{a}$ are the MOs and MO energies of the entire system.

\begin{figure}
\includegraphics[width=0.25\textwidth]{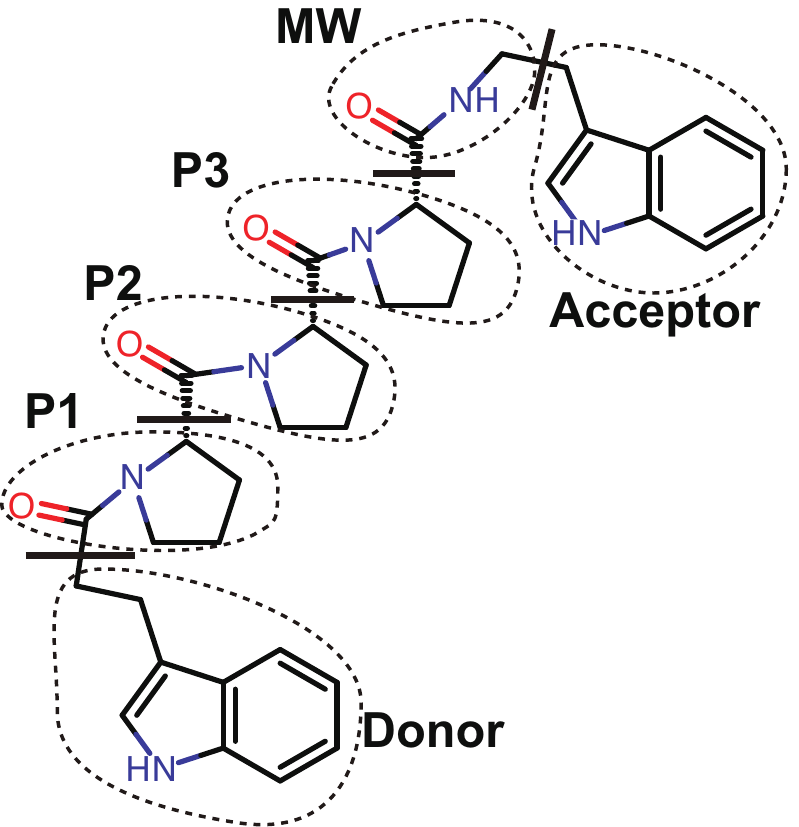}
\caption{
    Fragments in polyproline linker. P1-3 are three proline residues and MW denotes the main chain of Trp (W).
}
\end{figure}

We consider a \lq\lq minimal-valence
plus core\rq\rq\ rFMO space
that includes the same number of MOs as
the minimal basis set such as STO-3G.
We denote this LC(VC)MO space.
Other choices of
rFMO space,
LUMO, LUMO+2, LUMO+6, LUMO+10, 
to check the dependence on the size of rFMO space,
are obtained by augmenting
the lower unoccupied MOs of each fragment to the occupied space. 

The 6-31G(d) basis set
was used. The total number of basis AOs for the entire system is 774. 
The FMO2 and FMO3 calculations include additional 50 AOs from the BDAs.
The number of FMOs in the occupied,
LC(VC)MO, and full spaces for each fragment 
are summarized in Table I.
For basis sets larger
than the double-zeta basis, the LC(VC)MO scheme gives significant
reduction.
Moreover, 
it removed
the problem of linear-dependence observed previously \cite{Kobori2013},
as the smallest eigenvalue of the overlap matrix in the LC(VC)MO space was 0.225.

We used the program GAMESS \cite{Schmidt1993} 
for the conventional FMO calculation \cite{Fedorov2004}.
To estimate all inter-fragment tunneling currents 
including the long-distance ones,
we did not employ the electrostatic dimer (ES-DIM) approximation \cite{Nakano2002}
that avoids self-consistent field calculations of
the far separated dimers in FMO calculations.

\begin{table}
Table I. Number of MOs in the rFMO spaces.
\\
\baselineskip 12pt
\begin{tabular}{|c|c|c|c|c|c|c|c|}
\hline 
 & D & P1 & P2 & P3 & MW & A & Total \\
\hline 
\hline 
Occupied & 38 & 26 & 26 & 26 & 15 & 35 & 166 \\
\hline 
LC(VC)MO & 64 & 42 & 42 & 42 & 23 & 59 & 272$^{a}$ \\
\hline 
Full & 184 & 129 & 129 & 129 & 76 & 177 & 824 (774$^{b}$) \\
\hline 
\end{tabular}
\\
${}^{a}$Corresponds to the minimal set.
\\
${}^{b}$Without BDAs.
\end{table}

\section{Results}
\label{sec:results}

\subsection{Errors in MO energy}

\begin{table}
Table II.
MO energy gap and errors (in eV) of transPP helix from FMO-LC(VC)MO calculations.
\\
\baselineskip 12pt
\begin{tabular}{|c|c|c|c|c|c|}
\hline 
 & Gap$^{a}$ & \multicolumn{2}{c|}{MAE$^{b}$} & \multicolumn{2}{c|}{RMS$^{c}$}\tabularnewline
\hline 
  &  & Occ & Uoc & Occ & Uoc\tabularnewline
\hline 
FMO2 & 10.66 & 0.107 (\#15) & 22.4 (\#272) & 0.0345 & 3.92\tabularnewline
\hline 
FMO3 & 10.67 & 0.102 (\#27) & 14.6 (\#272) & 0.0266 & 3.13\tabularnewline
\hline 
FMO6 & 10.63 & 0.0934 (\#65) & 14.6 (\#272) & 0.0223 & 3.13\tabularnewline
\hline 
\end{tabular}
\\
$^{a}$HOMO-LUMO gap. The reference RHF value is 10.63 eV.
\\
$^{b}$Maximum absolute error of MO energies. 
Occ / Uoc denote occupied / unoccupied MOs.
In parentheses are the MO numbers that exhibit the MAE.
\\
$^{c}$Root-mean-square error of MO energies.
\end{table}

\begin{table}
Table III.
Same as Table II but for cisPP helix.
\\
\baselineskip 12pt
\begin{tabular}{|c|c|c|c|c|c|}
\hline 
 & Gap$^{a}$ & \multicolumn{2}{c|}{MAE$^{}$} & \multicolumn{2}{c|}{RMS$^{}$}\tabularnewline
\hline 
\hline 
 &  & Occ & Uoc & Occ & Uoc\tabularnewline
\hline 
FMO2 & 9.885 & 0.271 (\#19) & 22.18 (\#272) & 0.0504 & 3.74\tabularnewline
\hline 
FMO3 & 9.890 & 0.0988 (\#40) & 14.89 (\#272) & 0.0227 & 3.12\tabularnewline
\hline 
FMO6 & 9.909 & 0.0863 (\#65) & 14.95 (\#272) & 0.0213 & 3.14\tabularnewline
\hline 
\end{tabular}
\\
$^{a}$The reference RHF value is 9.904 eV.
\end{table}

The diagonalization of the FMO-LCMO Hamiltonian matrix
can provide approximate canonical MOs 
and corresponding energies for the
entire system \cite{Tsuneyuki2009, Kobori2013}.
First we assess accuracy of the FMO-LC(VC)MO method with regard 
to the MO energies.
The computed errors from the RHF calculation of the entire system
are summarized in Tables II and III
for transPP and cisPP, respectively.
For the occupied MOs, the maximum absolute error (MAE)
was observed at different MO numbers 
for FMO2, FMO3, and FMO6 calculations 
(for instance, Nos. 15, 27, and 65 for transPP)
but these commonly involve the BDA.
The MAE of unoccupied MOs was always observed at the highest MO (No. 272).
The root-mean-squares error is notably reduced from FMO2 to FMO3, 
but not so much from FMO3 to FMO6,
indicating nearly converged accuracy at the FMO3 level.

\subsection{ET coupling energy}
\label{subsec:coupling}

\begin{figure}
\includegraphics[width=0.35\textwidth]{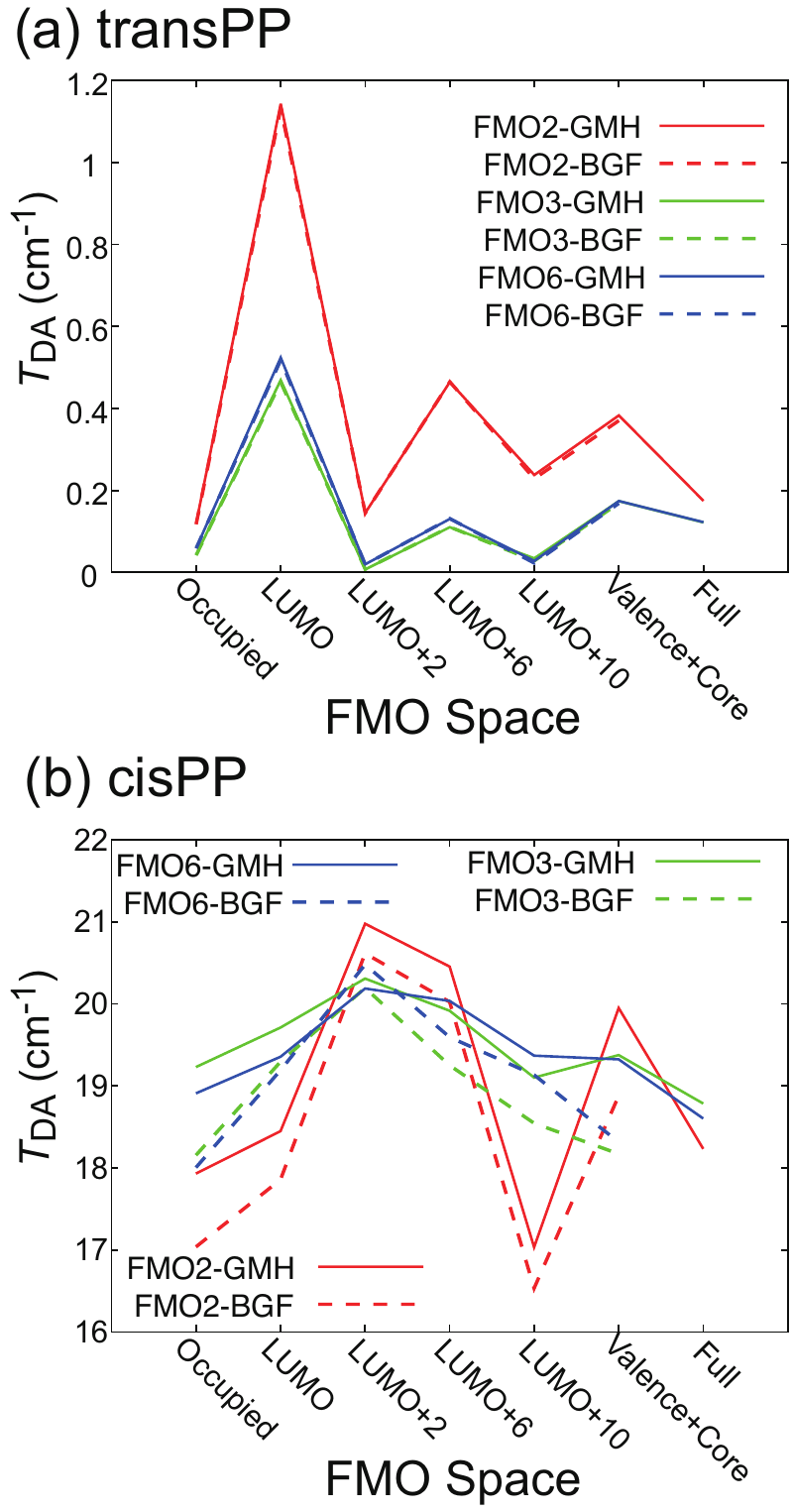}
\caption{
    Transfer matrix element $T_\textrm{DA}$ with various rFMO spaces for (a) transPP and (b) cisPP.
}
\end{figure}

\begin{figure}
\includegraphics[width=0.47\textwidth]{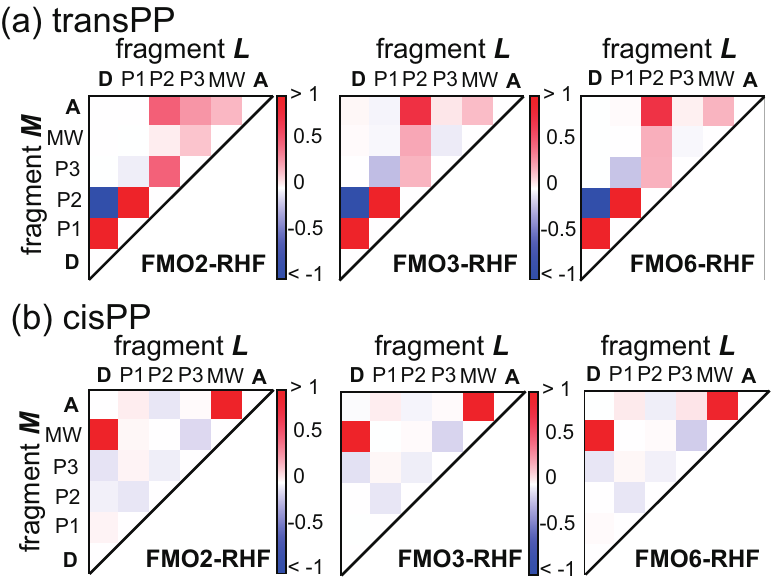}
\caption{
    Normalized inter-fragment tunneling current ${\cal K}_{L,M}$ from fragment $L$ to $M$
    with FMO2-, FMO3-, and FMO6-LC(VC)MO methods for (a) transPP and (b) cisPP.
}
\end{figure}

\begin{table}
Table IV.
Transfer matrix element $T_\textrm{DA}$ (in cm$^{-1}$) with various rFMO spaces 
for transPP complex.
\\
\baselineskip 12pt
\begin{tabular}{|c|c|c|c|c|c|c|}
\hline 
FMO space & \multicolumn{2}{c|}{FMO2} & \multicolumn{2}{c|}{FMO3} & \multicolumn{2}{c|}{FMO6}\tabularnewline
\hline 
         &  GMH   &  BGF   &  GMH    &  BGF    &  GMH    &  BGF    \tabularnewline
\hline 
Full     & 0.1745 &  ---   & 0.1219  &  ---    & 0.1224  &  ---    \tabularnewline
\hline 
LC(VC)MO & 0.3826 & 0.3702 & 0.1738  & 0.1688  & 0.1740  & 0.1687  \tabularnewline
\hline 
LUMO+10  & 0.2375 & 0.2286 & 0.03438 & 0.02850 & 0.02806 & 0.02301 \tabularnewline
\hline 
LUMO+6   & 0.4642 & 0.4646 & 0.1101  & 0.1108  & 0.1303  & 0.1312  \tabularnewline
\hline 
LUMO+2   & 0.1451 & 0.1448 & 0.07364 & 0.07956 & 0.0200  & 0.0195  \tabularnewline
\hline 
LUMO     & 1.143  & 1.132  & 0.4641  & 0.4638  & 0.5238  & 0.5176  \tabularnewline
\hline 
Occupied & 0.1198 & 0.1173 & 0.04252 & 0.04199 & 0.05932 & 0.05858 \tabularnewline
\hline 
\end{tabular}
\end{table}

\begin{table}
Table V.
Same as Table IV but for cisPP complex.
\\
\baselineskip 12pt
\begin{tabular}{|c|c|c|c|c|c|c|}
\hline 
& \multicolumn{2}{c|}{FMO2} & \multicolumn{2}{c|}{FMO3} & \multicolumn{2}{c|}{FMO6}\tabularnewline
\hline 
         &  GMH   &  BGF   &  GMH   &  BGF   &  GMH   &  BGF   \tabularnewline
\hline 
Full     &  18.23 &  ---   &  18.79 &  ---   &  18.60 &  ---   \tabularnewline
\hline 
LC(VC)MO &  19.95 &  18.88 &  19.38 &  18.17 &  19.32 &  18.31 \tabularnewline
\hline 
LUMO+10  &  17.03 &  16.53 &  19.10 &  18.54 &  19.37 &  19.13 \tabularnewline
\hline 
LUMO+6   &  22.45 &  20.02 &  19.92 &  19.25 &  20.04 &  19.59 \tabularnewline
\hline 
LUMO+2   &  29.98 &  20.61 &  20.31 &  20.19 &  20.19 &  20.48 \tabularnewline
\hline 
LUMO     &  18.45 &  17.86 &  19.71 &  19.29 &  19.36 &  19.19 \tabularnewline 
\hline 
Occupied &  17.93 &  17.04 &  19.23 &  18.16 &  18.91 &  18.01 \tabularnewline
\hline 
\end{tabular}
\end{table}

Next we examine the ET coupling matrix element $T_\textrm{DA}$. 
Figure 3 displays the computed $T_\textrm{DA}$ 
with varying rFMO spaces, from the \lq\lq occupied-only\rq\rq\ 
to the minimal-valence plus core (VC). 
The numerical values are listed in Tables IV and V.
The absolute value of $T_\textrm{DA}$ is about 20
times larger for \cPP\ than \tPP\ because of the shorter donor-acceptor distance in
the former. 
As seen in Fig. 3(a), the $T_\textrm{DA}$ in \tPP\ converge to the value of full space 
with an oscillation.
The behavior for \cPP\ in Fig. 3(b) is less simple;
the results of FMO2 exhibit notable oscillation
which is less prominent in FMO3 and FMO6.
For both \cPP\ and \tPP, 
FMO3 notably improves the $T_\textrm{DA}$ value over FMO2
and has been almost converged to FMO6.

Interestingly, the results with the occupied space appear closest to those with the full FMO space.
We consider this happend 
as the hole transfer is the principal mechanism for the
present system.
Thus, the addition of small number of LUMOs could have caused imbalance of
description.
However, generality of this view should be examined with more cases.

\subsection{ET pathway analysis}

Now we examine the ET pathway.
Figure 4(a) displays the normalized inter-fragment tunneling currents in \tPP,
comparing FMO2 and FMO3 with the reference FMO6.
In this figure, the LC(VC)MO space was employed.
The numerical values are listed in Table S8-S10 of the Supplementary Material.
The figure clearly indicates improvement of accuracy with FMO3 over FMO2.
The main pathway
is the forward ET of D$\rightarrow$P1$\rightarrow$P2$\rightarrow$A,
with a bifurcate back-flow of P2$\rightarrow$D.
The back-flow is due to destructive interference. 
(See Eq. (\ref{eq:tda}))
The figure indicates that FMO3 and FMO6 exhibit larger back-flow
than FMO2,
which explains the overestimate of $T_\textrm{DA}$ by FMO2
seen in Fig. 3(a).

The corresponding results for \cPP\ are displayed in Fig. 4(b). 
The numerical values are listed in Table S11-S13 of the Supplementary Material.
The major pathway is the forward flow of D$\rightarrow$MW$\rightarrow$A.
In contrast with \tPP, both FMO2 and FMO3 qualitatively reproduce
the pathway of reference FMO6 calculation.
This implies that, as noted in Sec. \ref{subsec:coupling}, 
the shorter donor-acceptor distance in \cPP\ makes the direct pathway
dominant, which could have masked the error stemming from the BDAs.

\section{Concluding Remarks}
\label{sec:concl}

Comparison between \cPP\ and \tPP\ indicated that 
the value of $T_\textrm{DA}$ is affected notably
by the selection of rFMO space.
Generally, the three-body correction
of FMO3 markedly improved the $T_\textrm{DA}$ value.
The ET pathway analysis is also made robust by the FMO3 correction,
especially 
when the BDAs are
involved in the ET pathway.

We employed the restricted Hartree-Fock (RHF) wave function
in this first report.
To include electron correlation effects,
an efficient way would be with the density functional theory.
For instance, we have recently found that the Kohn-Sham orbitals from
the long-range corrected functional give accurate electronic coupling energies
with non-empirical tuning of the range-separation parameter \cite{Nishioka15}.
The FMO3 correction will also make this scheme versatile.
To the correlated wave functions such as
the configuration interaction and coupled-cluster, 
extension of the FMO-LCMO scheme 
seems less straightforward
but deserves further examination.

%

\section*{Acknowledgments}
The authors acknowledge support from KAKENHI No. 20108017 (``$\pi$-space'').
H. K.-N. also acknowledges support from Collaborative Research Program for Young Scientists
of ACCMS and IIMC, Kyoto University.

\onecolumngrid
\newpage

\section*{\Large Supplementary Material}

\section*{Optimized Structures}

\begin{center}
\includegraphics[width=0.5\textwidth]{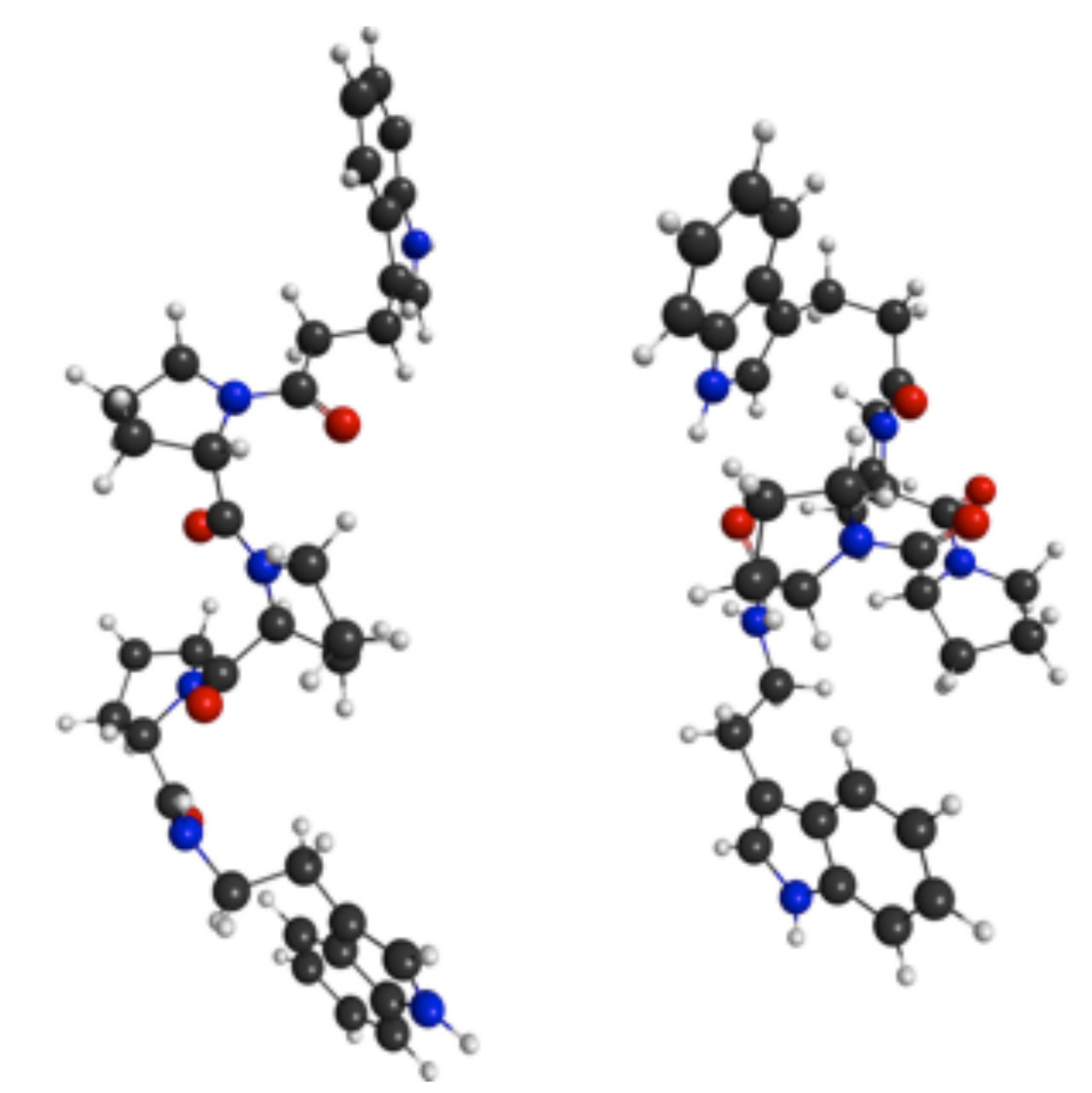}
\\
Fig.S1
Optimized structures of transPP (left) and cisPP (right)
at the B3LYP/6-31G(d)-D3 level.
\\
Cartesian coordinates are tabulated in Table S1 next page.
\end{center}

\newpage
\begin{center}
\textbf{Table S1.  } 
Cartesian coordinates (in \AA) of the structures in Fig.S1.
\\
\tiny
\begin{tabular}{|l|rrr|rrr|}
      \hline
      & \multicolumn{3}{c|}{transPP}             & \multicolumn{3}{c|}{cisPP} \\
      \hline
      C   &  -5.372434  &  -0.248416  &   0.062586  &  -4.971844  &   1.699604  &   0.475856  \\[-0.4em]
      H   &  -5.223344  &  -0.044113  &   1.132175  &  -5.246735  &   1.306885  &   1.457460  \\[-0.4em]
      C   &  -5.735621  &  -1.728813  &  -0.135761  &  -5.603116  &   0.816940  &  -0.639827  \\[-0.4em]
      H   &  -5.854165  &  -1.918929  &  -1.209884  &  -6.666165  &   0.676488  &  -0.407233  \\[-0.4em]
      H   &  -4.890183  &  -2.346251  &   0.183729  &  -5.569017  &   1.352615  &  -1.594752  \\[-0.4em]
      C   &  -6.981602  &  -2.109259  &   0.607296  &  -4.909142  &  -0.504208  &  -0.784947  \\[-0.4em]
      C   &  -7.071459  &  -2.893518  &   1.729218  &  -3.824857  &  -0.753603  &  -1.587351  \\[-0.4em]
      C   &  -8.326197  &  -1.681855  &   0.304662  &  -5.123114  &  -1.702775  &  -0.013530  \\[-0.4em]
      H   &  -6.288511  &  -3.406216  &   2.271782  &  -3.309798  &  -0.095425  &  -2.272372  \\[-0.4em]
      N   &  -8.389735  &  -2.978996  &   2.146910  &  -3.373190  &  -2.045405  &  -1.405494  \\[-0.4em]
      C   &  -9.184333  &  -2.246872  &   1.289035  &  -4.130589  &  -2.642560  &  -0.416482  \\[-0.4em]
      C   &  -8.887866  &  -0.883154  &  -0.706502  &  -6.043564  &  -2.069266  &   0.981646  \\[-0.4em]
      H   &  -8.719400  &  -3.517801  &   2.931645  &  -2.435580  &  -2.309688  &  -1.675676  \\[-0.4em]
      C   & -10.566610  &  -2.032309  &   1.287072  &  -4.041872  &  -3.917355  &   0.152543  \\[-0.4em]
      C   & -10.260725  &  -0.666079  &  -0.713117  &  -5.953903  &  -3.333529  &   1.553712  \\[-0.4em]
      H   &  -8.256084  &  -0.447998  &  -1.476867  &  -6.809389  &  -1.368331  &   1.305246  \\[-0.4em]
      H   & -11.207086  &  -2.472724  &   2.046967  &  -3.281526  &  -4.624352  &  -0.168891  \\[-0.4em]
      C   & -11.091445  &  -1.234430  &   0.275610  &  -4.960624  &  -4.247624  &   1.144668  \\[-0.4em]
      H   & -10.705317  &  -0.051496  &  -1.491208  &  -6.656798  &  -3.624582  &   2.329844  \\[-0.4em]
      H   & -12.161447  &  -1.047689  &   0.245609  &  -4.913849  &  -5.228879  &   1.609710  \\[-0.4em]
      H   &  -6.206356  &   0.395328  &  -0.245078  &  -5.349846  &   2.726995  &   0.398016  \\[-0.4em]
      C   &   5.537765  &  -0.621081  &  -0.911341  &   3.744816  &  -1.563074  &  -2.100025  \\[-0.4em]
      H   &   4.971085  &  -0.715441  &   0.024180  &   3.283052  &  -2.318982  &  -1.451355  \\[-0.4em]
      H   &   4.835236  &  -0.809198  &  -1.733596  &   3.749375  &  -1.983068  &  -3.112937  \\[-0.4em]
      C   &   6.657992  &  -1.613750  &  -0.915526  &   5.134565  &  -1.259284  &  -1.628448  \\[-0.4em]
      C   &   7.020002  &  -2.495988  &  -1.901059  &   6.241302  &  -0.985179  &  -2.393681  \\[-0.4em]
      C   &   7.604525  &  -1.790364  &   0.159501  &   5.535237  &  -1.082933  &  -0.253496  \\[-0.4em]
      H   &   6.565291  &  -2.676172  &  -2.866018  &   6.356238  &  -1.003598  &  -3.469127  \\[-0.4em]
      N   &   8.136541  &  -3.215375  &  -1.503844  &   7.305888  &  -0.646955  &  -1.578659  \\[-0.4em]
      C   &   8.515820  &  -2.804473  &  -0.240758  &   6.903724  &  -0.697270  &  -0.258887  \\[-0.4em]
      C   &   7.753870  &  -1.183072  &   1.419283  &   4.875484  &  -1.224726  &   0.980385  \\[-0.4em]
      H   &   8.586623  &  -3.937864  &  -2.042346  &   8.237522  &  -0.431243  &  -1.896542  \\[-0.4em]
      C   &   9.565340  &  -3.227275  &   0.582218  &   7.611976  &  -0.439028  &   0.919660  \\[-0.4em]
      C   &   8.796993  &  -1.601414  &   2.237659  &   5.572622  &  -0.966457  &   2.154456  \\[-0.4em]
      H   &   7.068708  &  -0.399383  &   1.735028  &   3.839319  &  -1.549788  &   1.018992  \\[-0.4em]
      H   &  10.255656  &  -4.004976  &   0.265163  &   8.657608  &  -0.143881  &   0.896502  \\[-0.4em]
      C   &   9.691445  &  -2.611843  &   1.823798  &   6.927450  &  -0.573215  &   2.122706  \\[-0.4em]
      H   &   8.929918  &  -1.143033  &   3.213926  &   5.071528  &  -1.072891  &   3.112393  \\[-0.4em]
      H   &  10.496578  &  -2.915789  &   2.487619  &   7.447318  &  -0.377366  &   3.056129  \\[-0.4em]
      C   &  -4.107769  &   0.141384  &  -0.692792  &  -3.449532  &   1.661316  &   0.436860  \\[-0.4em]
      O   &  -3.389861  &  -0.693300  &  -1.249382  &  -2.802119  &   1.024540  &   1.265214  \\[-0.4em]
      N   &  -3.801938  &   1.468648  &  -0.723001  &  -2.822496  &   2.292208  &  -0.603441  \\[-0.4em]
      C   &  -4.519514  &   2.557257  &  -0.047672  &  -3.397580  &   3.231957  &  -1.573137  \\[-0.4em]
      H   &  -4.764988  &   2.287502  &   0.984086  &  -3.932046  &   4.043672  &  -1.069119  \\[-0.4em]
      H   &  -5.460105  &   2.778722  &  -0.572427  &  -4.102053  &   2.716797  &  -2.237671  \\[-0.4em]
      C   &  -3.538067  &   3.737284  &  -0.122149  &  -2.174605  &   3.756137  &  -2.354794  \\[-0.4em]
      H   &  -4.052467  &   4.702717  &  -0.143263  &  -1.834151  &   4.693190  &  -1.908457  \\[-0.4em]
      H   &  -2.865474  &   3.705951  &   0.737251  &  -2.402315  &   3.931080  &  -3.410844  \\[-0.4em]
      C   &  -2.732507  &   3.454760  &  -1.402036  &  -1.104717  &   2.669638  &  -2.140698  \\[-0.4em]
      H   &  -1.764747  &   3.963334  &  -1.416004  &  -0.083775  &   3.037010  &  -2.287694  \\[-0.4em]
      H   &  -3.293969  &   3.765528  &  -2.291050  &  -1.262660  &   1.817426  &  -2.813270  \\[-0.4em]
      C   &  -2.585924  &   1.918316  &  -1.410697  &  -1.370078  &   2.210431  &  -0.693736  \\[-0.4em]
      H   &  -2.575908  &   1.509928  &  -2.421824  &  -1.071216  &   1.177799  &  -0.518523  \\[-0.4em]
      C   &  -1.347668  &   1.475309  &  -0.612773  &  -0.719984  &   3.190884  &   0.306623  \\[-0.4em]
      O   &  -1.211397  &   1.849419  &   0.559113  &  -1.248364  &   4.274252  &   0.548583  \\[-0.4em]
      N   &  -0.425028  &   0.704236  &  -1.232154  &   0.495515  &   2.857868  &   0.824329  \\[-0.4em]
      C   &  -0.518913  &   0.088221  &  -2.570280  &   1.140211  &   3.740198  &   1.810419  \\[-0.4em]
      H   &  -0.646138  &   0.853719  &  -3.341947  &   0.401122  &   4.081791  &   2.539089  \\[-0.4em]
      H   &  -1.380965  &  -0.585728  &  -2.588841  &   1.540597  &   4.625377  &   1.300008  \\[-0.4em]
      C   &   0.818137  &  -0.652447  &  -2.724359  &   2.244349  &   2.866828  &   2.426730  \\[-0.4em]
      H   &   1.566848  &   0.017207  &  -3.152519  &   1.859688  &   2.365842  &   3.317980  \\[-0.4em]
      H   &   0.721278  &  -1.536875  &  -3.361159  &   3.132182  &   3.444693  &   2.700144  \\[-0.4em]
      C   &   1.212747  &  -0.997368  &  -1.279471  &   2.534078  &   1.818781  &   1.337580  \\[-0.4em]
      H   &   2.284841  &  -1.177909  &  -1.158622  &   3.019520  &   0.915849  &   1.719994  \\[-0.4em]
      H   &   0.674035  &  -1.886868  &  -0.934171  &   3.180702  &   2.238175  &   0.556884  \\[-0.4em]
      C   &   0.728401  &   0.219704  &  -0.462839  &   1.137358  &   1.545298  &   0.740190  \\[-0.4em]
      H   &   0.397879  &  -0.071293  &   0.533795  &   1.200129  &   1.236721  &  -0.305125  \\[-0.4em]
      C   &   1.814560  &   1.304093  &  -0.407202  &   0.350024  &   0.519602  &   1.585136  \\[-0.4em]
      O   &   2.311270  &   1.711646  &  -1.469698  &  -0.063050  &   0.816286  &   2.696782  \\[-0.4em]
      N   &   2.245189  &   1.734611  &   0.801642  &   0.109003  &  -0.701920  &   1.012846  \\[-0.4em]
      C   &   1.581920  &   1.574677  &   2.109101  &  -0.750283  &  -1.664897  &   1.725561  \\[-0.4em]
      H   &   2.073360  &   0.778692  &   2.683633  &  -1.771668  &  -1.277435  &   1.751920  \\[-0.4em]
      H   &   0.525249  &   1.351333  &   1.973216  &  -0.396251  &  -1.773778  &   2.757907  \\[-0.4em]
      C   &   1.805101  &   2.945368  &   2.756388  &  -0.590345  &  -2.955249  &   0.916101  \\[-0.4em]
      H   &   1.703337  &   2.913972  &   3.845414  &  -1.327142  &  -2.982617  &   0.113673  \\[-0.4em]
      H   &   1.060574  &   3.646428  &   2.361153  &  -0.721310  &  -3.850699  &   1.529849  \\[-0.4em]
      C   &   3.221796  &   3.325949  &   2.296659  &   0.825265  &  -2.842544  &   0.326309  \\[-0.4em]
      H   &   3.975112  &   2.859595  &   2.938198  &   0.983704  &  -3.473377  &  -0.553830  \\[-0.4em]
      H   &   3.396902  &   4.405422  &   2.291320  &   1.578495  &  -3.112561  &   1.076192  \\[-0.4em]
      C   &   3.348782  &   2.725557  &   0.885964  &   0.969153  &  -1.333235  &   0.007007  \\[-0.4em]
      H   &   3.202190  &   3.464481  &   0.091699  &   2.000997  &  -1.012468  &   0.151305  \\[-0.4em]
      C   &   4.686718  &   1.977699  &   0.671114  &   0.520111  &  -1.046805  &  -1.437526  \\[-0.4em]
      O   &   5.354509  &   1.553037  &   1.612600  &  -0.626555  &  -1.281195  &  -1.817467  \\[-0.4em]
      N   &   5.017290  &   1.802128  &  -0.636757  &   1.444859  &  -0.531755  &  -2.301209  \\[-0.4em]
      H   &   4.254915  &   1.953291  &  -1.295434  &   1.091646  &  -0.444695  &  -3.247304  \\[-0.4em]
      C   &   6.048153  &   0.837217  &  -0.993796  &   2.868322  &  -0.289928  &  -2.077600  \\[-0.4em]
      H   &   6.883713  &   0.972767  &  -0.304518  &   3.009560  &   0.233150  &  -1.127806  \\[-0.4em]
      H   &   6.406451  &   1.060567  &  -2.003641  &   3.208838  &   0.401611  &  -2.854122  \\
      \hline
\end{tabular}
\end{center}

\normalsize
\newpage
\section*{Electron transfer energy}

\subsection*{transPP complex}

\begin{center}
{Table S2.} MO energy gap, maximum absolute error (MAE), root-mean-squares (RMS) error, and transfer matrix element $T_\textrm{DA}$
with FMO2-LCMO method for transPP complex.
\\[0.2em]
\begin{tabular}{|c|c|c|c|c|c|c|c|}
\hline 
 & Gap (eV) & \multicolumn{2}{c|}{MAE (eV)} & \multicolumn{2}{c|}{RMS (eV)} & \multicolumn{2}{c|}{$T_\textrm{DA}$ (cm$^{-1}$)}\\
\hline 
\hline 
 &  & Occ & Uoc & Occ & Uoc & GMH & BGF\\
\hline 
Full & 10.66 & 0.195 & 0.218 & 0.0399 & 0.0508 & 0.1745 & ---\\
\hline 
LC(VC)MO & 10.66 & 0.107 & 22.4 & 0.0345 & 3.92 & 0.3826 & 0.3702\\
\hline 
LUMO+10 & 10.66 & 0.218 & 20.6 & 0.0479 & 4.53 & 0.2375 & 0.2286\\
\hline 
LUMO+6 & 10.66 & 0.231 & 12.9 & 0.0521 & 3.52 & 0.4642 & 0.4646\\
\hline 
LUMO+2 & 10.65 & 0.243 & 6.20 & 0.0563 & 2.01 & 0.1451 & 0.1448\\
\hline 
LUMO & 10.65 & 0.256 & 2.97 & 0.0600 & 1.22 & 1.143 & 1.132\\
\hline 
Occupied & --- & 0.257 & --- & 0.0620 & --- & 0.1198 & 0.1173\\
\hline 
\end{tabular}

\bigskip \bigskip
{Table S3.} Same as Table S2 but 
with FMO3-LCMO.
\\[0.2em]
\begin{tabular}{|c|c|c|c|c|c|c|c|}
\hline 
 & Gap (eV) & \multicolumn{2}{c|}{MAE (eV)} & \multicolumn{2}{c|}{RMS (eV)} & \multicolumn{2}{c|}{$T_\textrm{DA}$ (cm$^{-1}$)}\\
\hline 
\hline 
 &  & Occ & Uoc & Occ & Uoc & GMH & BGF\\
\hline 
Full & 10.67 & 0.0400 & 0.0293 & 0.0119 & 0.0080 & 0.1219 & ---\\
\hline 
LC(VC)MO & 10.67 & 0.102 & 14.6 & 0.0266 & 3.12 & 0.1738 & 0.1688\\
\hline 
LUMO+10 & 10.67 & 0.167 & 18.6 & 0.0383 & 4.27 & 0.03438 & 0.02850\\
\hline 
LUMO+6 & 10.67 & 0.179 & 13.2 & 0.0432 & 3.32 & 0.1101 & 0.1108\\
\hline 
LUMO+2 & 10.66 & 0.191 & 5.69 & 0.0479 & 1.87 & 0.07364 & 0.07956\\
\hline 
LUMO & 10.66 & 0.216 & 2.39 & 0.0515 & 1.10 & 0.4641 & 0.4638\\
\hline 
Occupied & --- & 0.232 & --- & 0.0534 & --- & 0.04252 & 0.04199\\
\hline 
\end{tabular}

\bigskip \bigskip
{Table S4.} Same as Table S2 but 
with FMO6-LCMO, i.e., for the entire Trp-(Pro)$_3$-Trp system.
\\[0.2em]
\begin{tabular}{|c|c|c|c|c|c|c|c|}
\hline 
 & Gap (eV) & \multicolumn{2}{c|}{MAE (eV)} & \multicolumn{2}{c|}{RMS (eV)} & \multicolumn{2}{c|}{$T_\textrm{DA}$ (cm$^{-1}$)}\\
\hline 
\hline 
 &  & Occ & Uoc & Occ & Uoc & GMH & BGF\\
\hline 
Full & 10.63 & --- & --- & --- & --- & 0.1224 & ---\\
\hline 
LC(VC)MO & 10.63 & 0.0934 & 14.6 & 0.0223 & 3.13 & 0.1740 & 0.1687\\
\hline 
LUMO+10 & 10.63 & 0.179 & 18.6 & 0.0352 & 4.27 & 0.02806 & 0.02301\\
\hline 
LUMO+6 & 10.63 & 0.192 & 13.2 & 0.0400 & 3.32 & 0.1303 & 0.1312\\
\hline 
LUMO+2 & 10.62 & 0.203 & 5.71 & 0.0451 & 1.87 & 0.0200 & 0.0195\\
\hline 
LUMO & 10.62 & 0.220 & 2.39 & 0.0489 & 1.09 & 0.5238 & 0.5176\\
\hline 
Occupied & --- & 0.235 & --- & 0.0508 & --- & 0.05932 & 0.05858\\
\hline 
\end{tabular}

\newpage
\subsection*{cisPP complex}

{Table S5.} Same as Table S2 but 
for cisPP complex.
\\[0.2em]
\begin{tabular}{|c|c|c|c|c|c|c|c|}
\hline 
 & Gap (eV) & \multicolumn{2}{c|}{MAE (eV)} & \multicolumn{2}{c|}{RMS (eV)} & \multicolumn{2}{c|}{$T_\textrm{DA}$ (cm$^{-1}$)}\\
\hline 
\hline 
 &  & Occ & Uoc & Occ & Uoc & GMH & BGF\\
\hline 
Full & 9.880 & 0.215 & 0.356 & 0.0571 & 0.0724 & 18.23 & ---\\
\hline 
LC(VC)MO & 9.885 & 0.271 & 22.2 & 0.0504 & 3.74 & 19.95 & 18.88\\
\hline 
LUMO+10 & 9.890 & 0.370 & 26.1 & 0.0610 & 4.93 & 17.03 & 16.53\\
\hline 
LUMO+6 & 9.894 & 0.405 & 15.0 & 0.0683 & 3.56 & 22.45 & 20.02\\
\hline 
LUMO+2 & 9.912 & 0.416 & 9.34 & 0.0712 & 2.86 & 20.98 & 20.61\\
\hline 
LUMO & 9.924 & 0.426 & 5.42 & 0.0752 & 2.01 & 18.45 & 17.86\\
\hline 
Occupied & --- & 0.443 & --- & 0.0782 & --- & 17.93 & 17.04\\
\hline 
\end{tabular}

\bigskip \bigskip
{Table S6.} Same as Table S5 but 
with FMO3-LCMO.
\\[0.2em]
\begin{tabular}{|c|c|c|c|c|c|c|c|}
\hline 
 & Gap (eV) & \multicolumn{2}{c|}{MAE (eV)} & \multicolumn{2}{c|}{RMS (eV)} & \multicolumn{2}{c|}{$T_\textrm{DA}$ (cm$^{-1}$)}\\
\hline 
\hline 
 &  & Occ & Uoc & Occ & Uoc & GMH & BGF\\
\hline 
Full & 9.886 & 0.0405 & 0.124 & 0.0096 & 0.0156 & 18.79 & ---\\
\hline 
LC(VC)MO & 9.890 & 0.0988 & 14.9 & 0.0227 & 3.12 & 19.38 & 18.17\\
\hline 
LUMO+10 & 9.900 & 0.172 & 21.3 & 0.0361 & 4.58 & 19.10 & 18.54\\
\hline 
LUMO+6 & 9.905 & 0.199 & 14.1 & 0.0449 & 3.30 & 19.92 & 19.25\\
\hline 
LUMO+2 & 9.926 & 0.224 & 7.92 & 0.0485 & 2.35 & 20.31 & 20.19\\
\hline 
LUMO & 9.935 & 0.278 & 4.38 & 0.0532 & 1.64 & 19.71 & 19.29\\
\hline 
Occupied & --- & 0.294 & --- & 0.0558 & --- & 19.23 & 18.16\\
\hline 
\end{tabular}

\bigskip \bigskip
{Table S7.} Same as Table S5 but 
with FMO6-LCMO. 
\\[0.2em]
\begin{tabular}{|c|c|c|c|c|c|c|c|}
\hline 
 & Gap (eV) & \multicolumn{2}{c|}{MAE (eV)} & \multicolumn{2}{c|}{RMS (eV)} & \multicolumn{2}{c|}{$T_\textrm{DA}$ (cm$^{-1}$)}\\
\hline 
\hline 
 &  & Occ & Uoc & Occ & Uoc & GMH & BGF\\
\hline 
Full & 9.909 & --- & --- & --- & --- & 18.60 & ---\\
\hline 
LC(VC)MO & 9.909 & 0.0863 & 14.9 & 0.0213 & 3.14 & 19.32 & 18.31\\
\hline 
LUMO+10 & 9.919 & 0.170 & 21.2 & 0.0361 & 4.58 & 19.37 & 19.13\\
\hline 
LUMO+6 & 9.924 & 0.208 & 14.0 & 0.0455 & 3.31 & 20.04 & 19.59\\
\hline 
LUMO+2 & 9.946 & 0.235 & 8.26 & 0.0493 & 2.41 & 20.19 & 20.48\\
\hline 
LUMO & 9.955 & 0.284 & 4.41 & 0.0541 & 1.66 & 19.36 & 19.19\\
\hline 
Occupied & --- & 0.307 & --- & 0.0567 & --- & 18.91 & 18.01\\
\hline 
\end{tabular}

\newpage
\section*{Normalized tunneling current}

\subsection*{transPP helix}

{Table S8.} Normalized tunneling current
in transPP 
with FMO2-LC(VC)MO.
\\[0.2em]
\begin{tabular}{|c|c|c|c|c|c|}
\hline 
 & D & P1 & P2 & P3 & MW\\
\hline 
\hline 
P1 & 2.012 &  &  &  & \\
\hline 
P2 & -1.011 & 2.054 &  &  & \\
\hline 
P3 & -0.002 & -0.042 & 0.492 &  & \\
\hline 
MW & 0.000 & 0.000 & 0.046 & 0.149 & \\
\hline 
A & 0.000 & 0.000 & 0.506 & 0.300 & 0.194\\
\hline 
\end{tabular}

\bigskip \bigskip
{Table S9.} Same as Table S8 but 
with FMO3-LC(VC)MO.
\\[0.2em]
\begin{tabular}{|c|c|c|c|c|c|}
\hline 
 & D & P1 & P2 & P3 & MW\\
\hline 
\hline 
P1 & 3.496 &  &  &  & \\
\hline 
P2 & -2.534 & 3.738 &  &  & \\
\hline 
P3 & 0.006 & -0.197 & 0.201 &  & \\
\hline 
MW & 0.014 & -0.020 & 0.237 & -0.053 & \\
\hline 
A & 0.018 & -0.024 & 0.766 & 0.062 & 0.178\\
\hline 
\end{tabular}

\bigskip \bigskip
{Table S10.} Same as Table S8 but 
with FMO6-LC(VC)MO. 
\\[0.2em]
\begin{tabular}{|c|c|c|c|c|c|}
\hline 
 & D & P1 & P2 & P3 & MW\\
\hline 
\hline 
P1 & 3.531 &  &  &  & \\
\hline 
P2 & -2.521 & 3.694 &  &  & \\
\hline 
P3 & -0.007 & -0.179 & 0.204 &  & \\
\hline 
MW & -0.002 & 0.006 & 0.213 & -0.018 & \\
\hline 
A & -0.001 & 0.010 & 0.755 & 0.037 & 0.198\\
\hline 
\end{tabular}


\newpage
\subsection*{cisPP helix}

{Table S11.} Same as Table S8 but 
for cisPP complex.
\\[0.2em]
\begin{tabular}{|c|c|c|c|c|c|}
\hline 
 & D & P1 & P2 & P3 & MW\\
\hline 
\hline 
P1 & 0.025 &  &  &  & \\
\hline 
P2 & -0.035 & -0.067 &  &  & \\
\hline 
P3 & -0.073 & 0.029 & -0.043 &  & \\
\hline 
MW & 1.082 & 0.021 & 0.007 & -0.104 & \\
\hline 
A & 0.002 & 0.043 & -0.066 & 0.015 & 1.006\\
\hline 
\end{tabular}

\bigskip \bigskip
{Table S12.} Same as Table S11 but 
with FMO3-LC(VC)MO.
\\[0.2em]
\begin{tabular}{|c|c|c|c|c|c|}
\hline 
 & D & P1 & P2 & P3 & MW\\
\hline 
\hline 
P1 & -0.003 &  &  &  & \\
\hline 
P2 & 0.003 & -0.064 &  &  & \\
\hline 
P3 & -0.084 & 0.019 & -0.045 &  & \\
\hline 
MW & 1.091 & -0.004 & 0.009 & -0.123 & \\
\hline 
A & -0.008 & 0.046 & -0.025 & 0.013 & 0.973\\
\hline 
\end{tabular}

\bigskip \bigskip
{Table S13.} Same as Table S11 but 
with FMO6-LC(VC)MO. 
\\[0.2em]
\begin{tabular}{|c|c|c|c|c|c|}
\hline 
 & D & P1 & P2 & P3 & MW\\
\hline 
\hline 
P1 & 0.009 &  &  &  & \\
\hline 
P2 & 0.001 & -0.067 &  &  & \\
\hline 
P3 & -0.067 & 0.023 & -0.035 &  & \\
\hline 
MW & 1.060 & 0.001 & 0.010 & -0.142 & \\
\hline 
A & -0.003 & 0.052 & -0.040 & 0.063 & 0.928\\
\hline 
\end{tabular}


\end{center}


\begin{thebibliography}{18}
\expandafter\ifx\csname natexlab\endcsname\relax\def\natexlab#1{#1}\fi
\expandafter\ifx\csname bibnamefont\endcsname\relax
  \def\bibnamefont#1{#1}\fi
\expandafter\ifx\csname bibfnamefont\endcsname\relax
  \def\bibfnamefont#1{#1}\fi
\expandafter\ifx\csname citenamefont\endcsname\relax
  \def\citenamefont#1{#1}\fi
\expandafter\ifx\csname url\endcsname\relax
  \def\url#1{\texttt{#1}}\fi
\expandafter\ifx\csname urlprefix\endcsname\relax\def\urlprefix{URL }\fi
\providecommand{\bibinfo}[2]{#2}
\providecommand{\eprint}[2][]{\url{#2}}

\bibitem[{\citenamefont{Moser et~al.}(1992)\citenamefont{Moser, Keske, Warncke,
  Farid, and Dutton}}]{Moser1992}
\bibinfo{author}{\bibfnamefont{C.~C.} \bibnamefont{Moser}},
  \bibinfo{author}{\bibfnamefont{J.~M.} \bibnamefont{Keske}},
  \bibinfo{author}{\bibfnamefont{K.}~\bibnamefont{Warncke}},
  \bibinfo{author}{\bibfnamefont{R.~S.} \bibnamefont{Farid}}, \bibnamefont{and}
  \bibinfo{author}{\bibfnamefont{P.~L.} \bibnamefont{Dutton}},
  \bibinfo{journal}{Nature} \textbf{\bibinfo{volume}{355}},
  \bibinfo{pages}{796} (\bibinfo{year}{1992}).

\bibitem[{\citenamefont{Dutton and Mosser}(1994)}]{Dutton1994}
\bibinfo{author}{\bibfnamefont{P.~L.} \bibnamefont{Dutton}} \bibnamefont{and}
  \bibinfo{author}{\bibfnamefont{C.~C.} \bibnamefont{Mosser}},
  \bibinfo{journal}{Proc. Natl. Acad. Sci. USA} \textbf{\bibinfo{volume}{91}},
  \bibinfo{pages}{10247} (\bibinfo{year}{1994}).

\bibitem[{\citenamefont{Winkler et~al.}(1999)\citenamefont{Winkler, Di{\
  }Bilio, Farrow, Richards, and Gray}}]{Winkler1999}
\bibinfo{author}{\bibfnamefont{J.~R.} \bibnamefont{Winkler}},
  \bibinfo{author}{\bibfnamefont{A.~J.} \bibnamefont{Di{\ }Bilio}},
  \bibinfo{author}{\bibfnamefont{N.~A.} \bibnamefont{Farrow}},
  \bibinfo{author}{\bibfnamefont{J.~H.} \bibnamefont{Richards}},
  \bibnamefont{and} \bibinfo{author}{\bibfnamefont{H.~B.} \bibnamefont{Gray}},
  \bibinfo{journal}{Pure Appl. Chem.} \textbf{\bibinfo{volume}{71}},
  \bibinfo{pages}{1753} (\bibinfo{year}{1999}).

\bibitem[{\citenamefont{Gray and Winkler}(2005)}]{GrayWinkler2005}
\bibinfo{author}{\bibfnamefont{H.~B.} \bibnamefont{Gray}} \bibnamefont{and}
  \bibinfo{author}{\bibfnamefont{J.~R.} \bibnamefont{Winkler}},
  \bibinfo{journal}{Proc. Natl. Acad. Sci. USA.}
  \textbf{\bibinfo{volume}{102}}, \bibinfo{pages}{3534} (\bibinfo{year}{2005}).

\bibitem[{\citenamefont{Farver and Pecht}(2011)}]{FarverPecht2011}
\bibinfo{author}{\bibfnamefont{O.}~\bibnamefont{Farver}} \bibnamefont{and}
  \bibinfo{author}{\bibfnamefont{I.}~\bibnamefont{Pecht}},
  \bibinfo{journal}{Coord. Chem. Rev.} \textbf{\bibinfo{volume}{255}},
  \bibinfo{pages}{757} (\bibinfo{year}{2011}).

\bibitem[{\citenamefont{McConnell}(1961)}]{McConnell1961}
\bibinfo{author}{\bibfnamefont{H.~M.} \bibnamefont{McConnell}},
  \bibinfo{journal}{J. Chem. Phys.} \textbf{\bibinfo{volume}{35}},
  \bibinfo{pages}{508} (\bibinfo{year}{1961}).

\bibitem[{\citenamefont{Kitaura et~al.}(1999)\citenamefont{Kitaura, Ikeo,
  Asada, Nakano, and Uebayasi}}]{Kitaura1999}
\bibinfo{author}{\bibfnamefont{K.}~\bibnamefont{Kitaura}},
  \bibinfo{author}{\bibfnamefont{E.}~\bibnamefont{Ikeo}},
  \bibinfo{author}{\bibfnamefont{T.}~\bibnamefont{Asada}},
  \bibinfo{author}{\bibfnamefont{T.}~\bibnamefont{Nakano}}, \bibnamefont{and}
  \bibinfo{author}{\bibfnamefont{M.}~\bibnamefont{Uebayasi}},
  \bibinfo{journal}{Chem. Phys. Lett.} \textbf{\bibinfo{volume}{313}},
  \bibinfo{pages}{701} (\bibinfo{year}{1999}).

\bibitem[{\citenamefont{Fedorov and Kitaura}(2007)}]{Fedorov2007}
\bibinfo{author}{\bibfnamefont{D.~G.} \bibnamefont{Fedorov}} \bibnamefont{and}
  \bibinfo{author}{\bibfnamefont{K.}~\bibnamefont{Kitaura}},
  \bibinfo{journal}{J. Phys. Chem. A} \textbf{\bibinfo{volume}{111}},
  \bibinfo{pages}{6904} (\bibinfo{year}{2007}).

\bibitem[{\citenamefont{Tanaka et~al.}(2014)\citenamefont{Tanaka, Mochizuki,
  Komeiji, Okiyama, and Fukuzawa}}]{Tanaka2014}
\bibinfo{author}{\bibfnamefont{S.}~\bibnamefont{Tanaka}},
  \bibinfo{author}{\bibfnamefont{Y.}~\bibnamefont{Mochizuki}},
  \bibinfo{author}{\bibfnamefont{Y.}~\bibnamefont{Komeiji}},
  \bibinfo{author}{\bibfnamefont{Y.}~\bibnamefont{Okiyama}}, \bibnamefont{and}
  \bibinfo{author}{\bibfnamefont{K.}~\bibnamefont{Fukuzawa}},
  \bibinfo{journal}{Phys. Chem. Chem. Phys.} \textbf{\bibinfo{volume}{16}},
  \bibinfo{pages}{10310} (\bibinfo{year}{2014}).

\bibitem[{\citenamefont{Yang and Lee}(1995)}]{Yang95}
\bibinfo{author}{\bibfnamefont{W.}~\bibnamefont{Yang}} \bibnamefont{and}
  \bibinfo{author}{\bibfnamefont{T.~S.} \bibnamefont{Lee}},
  \bibinfo{journal}{J. Chem. Phys.} \textbf{\bibinfo{volume}{103}},
  \bibinfo{pages}{5674} (\bibinfo{year}{1995}).

\bibitem[{\citenamefont{Akama et~al.}(2009)\citenamefont{Akama, Kobayashi, and
  Nakai}}]{Akama2009}
\bibinfo{author}{\bibfnamefont{T.}~\bibnamefont{Akama}},
  \bibinfo{author}{\bibfnamefont{M.}~\bibnamefont{Kobayashi}},
  \bibnamefont{and} \bibinfo{author}{\bibfnamefont{H.}~\bibnamefont{Nakai}},
  \bibinfo{journal}{Int. J. Quant. Chem.} \textbf{\bibinfo{volume}{109}},
  \bibinfo{pages}{2706} (\bibinfo{year}{2009}).

\bibitem[{\citenamefont{Li and Piecuch}(2010)}]{Li2010}
\bibinfo{author}{\bibfnamefont{W.}~\bibnamefont{Li}} \bibnamefont{and}
  \bibinfo{author}{\bibfnamefont{P.}~\bibnamefont{Piecuch}},
  \bibinfo{journal}{J. Phys. Chem. A} \textbf{\bibinfo{volume}{114}},
  \bibinfo{pages}{6721} (\bibinfo{year}{2010}).

\bibitem[{\citenamefont{Aoki and Gu}(2012)}]{Aoki2012}
\bibinfo{author}{\bibfnamefont{Y.}~\bibnamefont{Aoki}} \bibnamefont{and}
  \bibinfo{author}{\bibfnamefont{F.~L.} \bibnamefont{Gu}},
  \bibinfo{journal}{Phys. Chem. Chem. Phys.} \textbf{\bibinfo{volume}{14}},
  \bibinfo{pages}{7640} (\bibinfo{year}{2012}).

\bibitem[{\citenamefont{Gordon et~al.}(2012)\citenamefont{Gordon, Fedorov,
  Pruitt, and Slipchenko}}]{Gordon2012}
\bibinfo{author}{\bibfnamefont{M.~S.} \bibnamefont{Gordon}},
  \bibinfo{author}{\bibfnamefont{D.~G.} \bibnamefont{Fedorov}},
  \bibinfo{author}{\bibfnamefont{S.~R.} \bibnamefont{Pruitt}},
  \bibnamefont{and} \bibinfo{author}{\bibfnamefont{L.~V.}
  \bibnamefont{Slipchenko}}, \bibinfo{journal}{Chem. Rev.}
  \textbf{\bibinfo{volume}{112}}, \bibinfo{pages}{632} (\bibinfo{year}{2012}).

\bibitem[{\citenamefont{Raghavachari and Saha}(2015)}]{Raghavachari2015}
\bibinfo{author}{\bibfnamefont{K.}~\bibnamefont{Raghavachari}}
  \bibnamefont{and} \bibinfo{author}{\bibfnamefont{A.}~\bibnamefont{Saha}},
  \bibinfo{journal}{Chem. Rev.} \textbf{\bibinfo{volume}{115}},
  \bibinfo{pages}{5643} (\bibinfo{year}{2015}).

\bibitem[{\citenamefont{Nishioka and Ando}(2011)}]{Nishioka2011}
\bibinfo{author}{\bibfnamefont{H.}~\bibnamefont{Nishioka}} \bibnamefont{and}
  \bibinfo{author}{\bibfnamefont{K.}~\bibnamefont{Ando}}, \bibinfo{journal}{J.
  Chem. Phys.} \textbf{\bibinfo{volume}{134}}, \bibinfo{pages}{204109}
  (\bibinfo{year}{2011}).

\bibitem[{\citenamefont{Kitoh-Nishioka and Ando}(2012)}]{Kitoh-Nishioka2012}
\bibinfo{author}{\bibfnamefont{H.}~\bibnamefont{Kitoh-Nishioka}}
  \bibnamefont{and} \bibinfo{author}{\bibfnamefont{K.}~\bibnamefont{Ando}},
  \bibinfo{journal}{J. Phys. Chem. B} \textbf{\bibinfo{volume}{116}},
  \bibinfo{pages}{12933} (\bibinfo{year}{2012}).

\bibitem[{\citenamefont{Kitoh-Nishioka and Ando}(2015)}]{Nishioka15}
\bibinfo{author}{\bibfnamefont{H.}~\bibnamefont{Kitoh-Nishioka}}
  \bibnamefont{and} \bibinfo{author}{\bibfnamefont{K.}~\bibnamefont{Ando}},
  \bibinfo{journal}{Chem. Phys. Lett.} \textbf{\bibinfo{volume}{621}},
  \bibinfo{pages}{96} (\bibinfo{year}{2015}).

\bibitem[{\citenamefont{Tsuneyuki et~al.}(2009)\citenamefont{Tsuneyuki, Kobori,
  Akagi, Sodeyama, Terakura, and Fukuyama}}]{Tsuneyuki2009}
\bibinfo{author}{\bibfnamefont{S.}~\bibnamefont{Tsuneyuki}},
  \bibinfo{author}{\bibfnamefont{T.}~\bibnamefont{Kobori}},
  \bibinfo{author}{\bibfnamefont{K.}~\bibnamefont{Akagi}},
  \bibinfo{author}{\bibfnamefont{K.}~\bibnamefont{Sodeyama}},
  \bibinfo{author}{\bibfnamefont{K.}~\bibnamefont{Terakura}}, \bibnamefont{and}
  \bibinfo{author}{\bibfnamefont{H.}~\bibnamefont{Fukuyama}},
  \bibinfo{journal}{Chem. Phys. Lett.} \textbf{\bibinfo{volume}{476}},
  \bibinfo{pages}{104} (\bibinfo{year}{2009}).

\bibitem[{\citenamefont{Kobori et~al.}(2013)\citenamefont{Kobori, Sodeyama,
  Otsuka, Tateyama, and Tsuneyuki}}]{Kobori2013}
\bibinfo{author}{\bibfnamefont{T.}~\bibnamefont{Kobori}},
  \bibinfo{author}{\bibfnamefont{K.}~\bibnamefont{Sodeyama}},
  \bibinfo{author}{\bibfnamefont{T.}~\bibnamefont{Otsuka}},
  \bibinfo{author}{\bibfnamefont{Y.}~\bibnamefont{Tateyama}}, \bibnamefont{and}
  \bibinfo{author}{\bibfnamefont{S.}~\bibnamefont{Tsuneyuki}},
  \bibinfo{journal}{J. Chem. Phys.} \textbf{\bibinfo{volume}{139}},
  \bibinfo{pages}{094113} (\bibinfo{year}{2013}).

\bibitem[{\citenamefont{Szabo and Ostlund}(1996)}]{SzaboOstlund}
\bibinfo{author}{\bibfnamefont{A.}~\bibnamefont{Szabo}} \bibnamefont{and}
  \bibinfo{author}{\bibfnamefont{N.~S.} \bibnamefont{Ostlund}},
  \emph{\bibinfo{title}{Modern Quantum Chemistry}} (\bibinfo{publisher}{Dover},
  \bibinfo{address}{New York}, \bibinfo{year}{1996}).

\bibitem[{\citenamefont{Isied et~al.}(1992)\citenamefont{Isied, Ogawa, 
and Wishart}}]{Isied1992}
\bibinfo{author}{\bibfnamefont{S.~S.}~\bibnamefont{Isied}},
  \bibinfo{author}{\bibfnamefont{M.~Y.}~\bibnamefont{Ogawa}},
  \bibnamefont{and}
  \bibinfo{author}{\bibfnamefont{J.~F.}~\bibnamefont{Wishart}},
  \bibinfo{journal}{Chem. Rev.} \textbf{\bibinfo{volume}{92}},
  \bibinfo{pages}{381} (\bibinfo{year}{1992}).

\bibitem[{\citenamefont{Ogawa et~al.}(1993)\citenamefont{Ogawa, Wishart,
  Young, Miller, and Isied}}]{Ogawa1993A}
\bibinfo{author}{\bibfnamefont{M.~Y.}~\bibnamefont{Ogawa}},
  \bibinfo{author}{\bibfnamefont{J.~F.}~\bibnamefont{Wishart}},
  \bibinfo{author}{\bibfnamefont{Z.}~\bibnamefont{Young}},
  \bibinfo{author}{\bibfnamefont{J.~R.}~\bibnamefont{Miller}}, \bibnamefont{and}
  \bibinfo{author}{\bibfnamefont{S.~S.}~\bibnamefont{Isied}},
  \bibinfo{journal}{J. Phys. Chem.} \textbf{\bibinfo{volume}{97}},
  \bibinfo{pages}{11456} (\bibinfo{year}{1993}).

\bibitem[{\citenamefont{Ogawa et~al.}(1993)\citenamefont{Ogawa, Moreira,
  Wishart, and Isied}}]{Ogawa1993B}
\bibinfo{author}{\bibfnamefont{M.~Y.}~\bibnamefont{Ogawa}},
  \bibinfo{author}{\bibfnamefont{I.}~\bibnamefont{Moreira}},
  \bibinfo{author}{\bibfnamefont{J.~F.}~\bibnamefont{Wishart}},
  \bibnamefont{and}
  \bibinfo{author}{\bibfnamefont{S.~S.}~\bibnamefont{Isied}},
  \bibinfo{journal}{Chem. Phys.} \textbf{\bibinfo{volume}{176}},
  \bibinfo{pages}{589} (\bibinfo{year}{1993}).

\bibitem[{\citenamefont{Wallrapp et~al.}(2009)\citenamefont{Wallrapp,
  Voityuk, and Guallar}}]{Wallrapp2009}
\bibinfo{author}{\bibfnamefont{F.}~\bibnamefont{Wallrapp}},
  \bibinfo{author}{\bibfnamefont{A.~A.}~\bibnamefont{Voityuk}},
  \bibnamefont{and}
  \bibinfo{author}{\bibfnamefont{V.}~\bibnamefont{Guallar}},
  \bibinfo{journal}{J. Chem. Theory Comput.} \textbf{\bibinfo{volume}{5}},
  \bibinfo{pages}{3312} (\bibinfo{year}{2009}).

\bibitem[{\citenamefont{Wallrapp et~al.}(2010)\citenamefont{Wallrapp,
  Voityuk, and Guallar}}]{Wallrapp2010}
\bibinfo{author}{\bibfnamefont{F.}~\bibnamefont{Wallrapp}},
  \bibinfo{author}{\bibfnamefont{A.~A.}~\bibnamefont{Voityuk}},
  \bibnamefont{and}
  \bibinfo{author}{\bibfnamefont{V.}~\bibnamefont{Guallar}},
  \bibinfo{journal}{J. Chem. Theory Comput.} \textbf{\bibinfo{volume}{6}},
  \bibinfo{pages}{3241} (\bibinfo{year}{2010}).

\bibitem[{\citenamefont{Cave and Newton}(1996)}]{Cave1996}
\bibinfo{author}{\bibfnamefont{R.~J.} \bibnamefont{Cave}} \bibnamefont{and}
  \bibinfo{author}{\bibfnamefont{M.~D.} \bibnamefont{Newton}},
  \bibinfo{journal}{Chem. Phys. Lett.} \textbf{\bibinfo{volume}{249}},
  \bibinfo{pages}{15} (\bibinfo{year}{1996}).

\bibitem[{\citenamefont{Skourtis and Beratan}(1999)}]{Skourtis1999}
\bibinfo{author}{\bibfnamefont{S.~S.} \bibnamefont{Skourtis}} \bibnamefont{and}
  \bibinfo{author}{\bibfnamefont{D.~N.} \bibnamefont{Beratan}},
  \bibinfo{journal}{Adv. Chem. Phys.} \textbf{\bibinfo{volume}{106}},
  \bibinfo{pages}{377} (\bibinfo{year}{1999}).

\bibitem[{\citenamefont{Regan and Onuchic}(1999)}]{Regan1999}
\bibinfo{author}{\bibfnamefont{J.~J.} \bibnamefont{Regan}} \bibnamefont{and}
  \bibinfo{author}{\bibfnamefont{J.~N.} \bibnamefont{Onuchic}},
  \bibinfo{journal}{Adv. Chem. Phys.} \textbf{\bibinfo{volume}{107}},
  \bibinfo{pages}{497} (\bibinfo{year}{1999}).

\bibitem[{\citenamefont{Stuchebrukhov}(2003)}]{Stuchebrukhov2003}
\bibinfo{author}{\bibfnamefont{A.~A.} \bibnamefont{Stuchebrukhov}},
  \bibinfo{journal}{Theor. Chem. Acc.} \textbf{\bibinfo{volume}{110}},
  \bibinfo{pages}{291} (\bibinfo{year}{2003}).

\bibitem[{\citenamefont{Stuchebrukhov}(1996)}]{Stuchebrukhov1996}
\bibinfo{author}{\bibfnamefont{A.~A.} \bibnamefont{Stuchebrukhov}},
  \bibinfo{journal}{J. Chem. Phys.} \textbf{\bibinfo{volume}{105}},
  \bibinfo{pages}{10819} (\bibinfo{year}{1996}).

\bibitem[{\citenamefont{Grimme et~al.}(2010)\citenamefont{Grimme, Antony,
  Ehrlich, and Krieg}}]{Grimme2010}
\bibinfo{author}{\bibfnamefont{S.}~\bibnamefont{Grimme}},
  \bibinfo{author}{\bibfnamefont{J.}~\bibnamefont{Antony}},
  \bibinfo{author}{\bibfnamefont{S.}~\bibnamefont{Ehrlich}}, \bibnamefont{and}
  \bibinfo{author}{\bibfnamefont{H.}~\bibnamefont{Krieg}},
  \bibinfo{journal}{J. Chem. Phys.} \textbf{\bibinfo{volume}{132}},
  \bibinfo{pages}{154104} (\bibinfo{year}{2010}).

\bibitem[{\citenamefont{Frisch et~al.}(2013)\citenamefont{Frisch, Trucks, Schlegel,
  Scuseria, Robb, Cheeseman, Scalmani, Barone, Mennucci, Petersson
  et~al.}}]{Gaussian09}
\bibinfo{author}{\bibfnamefont{M.~J.} \bibnamefont{Frisch}},
  \bibinfo{author}{\bibfnamefont{G.~W.} \bibnamefont{Trucks}},
  \bibinfo{author}{\bibfnamefont{H.~B.} \bibnamefont{Schlegel}},
  \bibinfo{author}{\bibfnamefont{G.~E.} \bibnamefont{Scuseria}},
  \bibinfo{author}{\bibfnamefont{M.~A.} \bibnamefont{Robb}},
  \bibinfo{author}{\bibfnamefont{J.~R.} \bibnamefont{Cheeseman}},
  \bibinfo{author}{\bibfnamefont{G.}~\bibnamefont{Scalmani}},
  \bibinfo{author}{\bibfnamefont{V.}~\bibnamefont{Barone}},
  \bibinfo{author}{\bibfnamefont{B.}~\bibnamefont{Mennucci}},
  \bibinfo{author}{\bibfnamefont{G.~A.} \bibnamefont{Petersson}},
  \bibnamefont{et~al.}, \emph{\bibinfo{title}{Gaussian~09 {R}evision {D}.01}},
  \bibinfo{note}{Gaussian Inc. Wallingford CT 2013}.

\bibitem[{\citenamefont{Schmidt et~al.}(1993)\citenamefont{Schmidt, Baldridge,
  Boatz, Elbert, Gordon, Jensen, Koseki, Matsunaga, Nguyen, Sue et~al.}}]{Schmidt1993}
\bibinfo{author}{\bibfnamefont{M.~W.}~\bibnamefont{Schmidt}},
  \bibinfo{author}{\bibfnamefont{K.~K.}~\bibnamefont{Baldridge}},
  \bibinfo{author}{\bibfnamefont{J.~A.}~\bibnamefont{Boatz}},
  \bibinfo{author}{\bibfnamefont{S.~T.}~\bibnamefont{Elbert}},
  \bibinfo{author}{\bibfnamefont{M.~S.}~\bibnamefont{Gordon}}, 
  \bibinfo{author}{\bibfnamefont{J.~H.}~\bibnamefont{Jensen}}, 
  \bibinfo{author}{\bibfnamefont{S.}~\bibnamefont{Koseki}}, 
  \bibinfo{author}{\bibfnamefont{N.}~\bibnamefont{Matsunaga}}, 
  \bibinfo{author}{\bibfnamefont{K.~A.}~\bibnamefont{Nguyen}}, 
  \bibinfo{author}{\bibfnamefont{S.}~\bibnamefont{Su}}, 
  \bibnamefont{et~al.},
  \bibinfo{journal}{J. Comput. Chem.} \textbf{\bibinfo{volume}{14}},
  \bibinfo{pages}{1347} (\bibinfo{year}{1993}).

\bibitem[{\citenamefont{Fedorov and Kitaura}(2004)}]{Fedorov2004}
\bibinfo{author}{\bibfnamefont{D.~G.} \bibnamefont{Fedorov}} \bibnamefont{and}
  \bibinfo{author}{\bibfnamefont{K.}~\bibnamefont{Kitaura}},
  \bibinfo{journal}{J. Chem. Phys.} \textbf{\bibinfo{volume}{120}},
  \bibinfo{pages}{6832} (\bibinfo{year}{2004}).

\bibitem[{\citenamefont{Nakano et~al.}(2002)\citenamefont{Nakano, Kaminuma,
  Sato, Fukuzawa, Akiyama, Uebayashi and Kitaura}}]{Nakano2002}
\bibinfo{author}{\bibfnamefont{T.}~\bibnamefont{Nakano}},
  \bibinfo{author}{\bibfnamefont{T.}~\bibnamefont{Kaminuma}},
  \bibinfo{author}{\bibfnamefont{T.}~\bibnamefont{Sato}},
  \bibinfo{author}{\bibfnamefont{K.}~\bibnamefont{Fukuzawa}},
  \bibinfo{author}{\bibfnamefont{Y.}~\bibnamefont{Akiyama}},
  \bibinfo{author}{\bibfnamefont{M.}~\bibnamefont{Uebayasi}},
  \bibnamefont{and}
  \bibinfo{author}{\bibfnamefont{K.}~\bibnamefont{Kitaura}},
  \bibinfo{journal}{Chem. Phys. Lett.} \textbf{\bibinfo{volume}{351}},
  \bibinfo{pages}{475} (\bibinfo{year}{2002}).

\end{thebibliography}
\end{document}